\documentclass{article}
\usepackage{psfig}
\usepackage{latexsym}
\begin{document}
\newcommand {\ee}{\end{equation}}
\newcommand {\bea}{\begin{eqnarray}}
\newcommand {\eea}{\end{eqnarray}}
\newcommand {\nn}{\nonumber \\}
\newcommand {\Tr}{{\rm Tr\,}}
\newcommand {\tr}{{\rm tr\,}}
\newcommand {\e}{{\rm e}}
\newcommand {\etal}{{\it et al.}}
\newcommand {\m}{\mu}
\newcommand {\n}{\nu}
\newcommand {\pl}{\partial}
\newcommand {\p} {\phi}
\newcommand {\vp}{\varphi}
\newcommand {\vpc}{\varphi_c}
\newcommand {\al}{\alpha}
\newcommand {\be}{\beta}
\newcommand {\ga}{\gamma}
\newcommand {\Ga}{\Gamma}
\newcommand {\x}{\xi}
\newcommand {\ka}{\kappa}
\newcommand {\la}{\lambda}
\newcommand {\La}{\Lambda}
\newcommand {\si}{\sigma}
\newcommand {\Si}{\Sigma}
\newcommand {\Th}{\Theta}
\newcommand {\om}{\omega}
\newcommand {\Om}{\Omega}
\newcommand {\ep}{\epsilon}
\newcommand {\vep}{\varepsilon}
\newcommand {\na}{\nabla}
\newcommand {\del}  {\delta}
\newcommand {\Del}  {\Delta}
\newcommand {\mn}{{\mu\nu}}
\newcommand {\ls}   {{\lambda\sigma}}
\newcommand {\ab}   {{\alpha\beta}}
\newcommand {\gd}   {{\gamma\delta}}
\newcommand {\half}{ {\frac{1}{2}} }
\newcommand {\third}{ {\frac{1}{3}} }
\newcommand {\fourth} {\frac{1}{4} }
\newcommand {\sixth} {\frac{1}{6} }
\newcommand {\sqg} {\sqrt{g}}
\newcommand {\sqtwo} {\sqrt{2}}
\newcommand {\fg}  {\sqrt[4]{g}}
\newcommand {\invfg}  {\frac{1}{\sqrt[4]{g}}}
\newcommand {\sqZ} {\sqrt{Z}}
\newcommand {\sqk} {\sqrt{\kappa}}
\newcommand {\sqt} {\sqrt{t}}
\newcommand {\sql} {\sqrt{l}}
\newcommand {\reg} {\frac{1}{\epsilon}}
\newcommand {\fpisq} {(4\pi)^2}
\newcommand {\Lcal}{{\cal L}}
\newcommand {\Ocal}{{\cal O}}
\newcommand {\Dcal}{{\cal D}}
\newcommand {\Ncal}{{\cal N}}
\newcommand {\Mcal}{{\cal M}}
\newcommand {\scal}{{\cal s}}
\newcommand {\Dvec}{{\hat D}}   
\newcommand {\dvec}{{\vec d}}
\newcommand {\Evec}{{\vec E}}
\newcommand {\Hvec}{{\vec H}}
\newcommand {\Vvec}{{\vec V}}
\newcommand {\rpl}{{\vec \partial}}
\def\overleftarrow#1{\vbox{\ialign{##\crcr
 $\leftarrow$\crcr\noalign{\kern-1pt\nointerlineskip}
 $\hfil\displaystyle{#1}\hfil$\crcr}}}
\def\lpl{{\overleftarrow\partial}}
\newcommand {\Btil}{{\tilde B}}
\newcommand {\ctil}{{\tilde c}}
\newcommand {\dtil}{{\tilde d}}
\newcommand {\Ftil}{{\tilde F}}
\newcommand {\Ktil}  {{\tilde K}}
\newcommand {\Ltil}  {{\tilde L}}
\newcommand {\mtil}{{\tilde m}}
\newcommand {\ttil} {{\tilde t}}
\newcommand {\Qtil}  {{\tilde Q}}
\newcommand {\Rtil}  {{\tilde R}}
\newcommand {\Stil}{{\tilde S}}
\newcommand {\Ztil}{{\tilde Z}}
\newcommand {\altil}{{\tilde \alpha}}
\newcommand {\betil}{{\tilde \beta}}
\newcommand {\deltil}{{\tilde \delta}}
\newcommand {\eptil}{{\tilde \epsilon}}
\newcommand {\etatil} {{\tilde \eta}}
\newcommand {\latil}{{\tilde \lambda}}
\newcommand {\Latil}{{\tilde \Lambda}}
\newcommand {\ptil}{{\tilde \phi}}
\newcommand {\Ptil}{{\tilde \Phi}}
\newcommand {\natil} {{\tilde \nabla}}
\newcommand {\xitil} {{\tilde \xi}}
\newcommand {\Ahat}{{\hat A}}
\newcommand {\ahat}{{\hat a}}
\newcommand {\Rhat}{{\hat R}}
\newcommand {\Shat}{{\hat S}}
\newcommand {\ehat}{{\hat e}}
\newcommand {\mhat}{{\hat m}}
\newcommand {\shat}{{\hat s}}
\newcommand {\Dhat}{{\hat D}}   
\newcommand {\Vhat}{{\hat V}}   
\newcommand {\xhat}{{\hat x}}
\newcommand {\Zhat}{{\hat Z}}
\newcommand {\Gahat}{{\hat \Gamma}}
\newcommand {\Phihat} {{\hat \Phi}}
\newcommand {\phihat} {{\hat \phi}}
\newcommand {\vphat} {{\hat \varphi}}
\newcommand {\nah} {{\hat \nabla}}
\newcommand {\etahat} {{\hat \eta}}
\newcommand {\omhat} {{\hat \omega}}
\newcommand {\psihat} {{\hat \psi}}
\newcommand {\thhat} {{\hat \theta}}
\newcommand {\gh}  {{\hat g}}
\newcommand {\abar}{{\bar a}}
\newcommand {\Abar}{{\bar A}}
\newcommand {\cbar}{{\bar c}}
\newcommand {\bbar}{{\bar b}}
\newcommand {\gbar}{\bar{g}}
\newcommand {\Bbar}{{\bar B}}
\newcommand {\fbar}{{\bar f}}
\newcommand {\Fbar}{{\bar F}}
\newcommand {\kbar}  {{\bar k}}
\newcommand {\Kbar}  {{\bar K}}
\newcommand {\Lbar}  {{\bar L}}
\newcommand {\Qbar}  {{\bar Q}}
\newcommand {\Zbar}  {{\bar Z}}
\newcommand {\albar}{{\bar \alpha}}
\newcommand {\bebar}{{\bar \beta}}
\newcommand {\labar}{{\bar \lambda}}
\newcommand {\psibar}{{\bar \psi}}
\newcommand {\vpbar}{{\bar \varphi}}
\newcommand {\Psibar}{{\bar \Psi}}
\newcommand {\chibar}{{\bar \chi}}
\newcommand {\sibar}{{\bar \sigma}}
\newcommand {\xibar}{{\bar \xi}}
\newcommand {\thbar}{{\bar \theta}}
\newcommand {\bbartil}{{\tilde {\bar b}}}
\newcommand {\aldot}{{\dot{\alpha}}}
\newcommand {\bedot}{{\dot{\beta}}}
\newcommand {\alp}{{\alpha'}}
\newcommand {\bep}{{\beta'}}
\newcommand {\gap}{{\gamma'}}
\newcommand {\bfZ} {{\bf Z}}
\newcommand {\BFd} {{\bf d}}
\newcommand  {\vz}{{v_0}}
\newcommand  {\ez}{{e_0}}
\newcommand  {\mz}{{m_0}}
\newcommand  {\xf}{{x^5}}
\newcommand  {\yf}{{y^5}}
\newcommand  {\Zt}{{Z$_2$}}
\newcommand {\intfx} {{\int d^4x}}
\newcommand {\intdX} {{\int d^5X}}
\newcommand {\inttx} {{\int d^2x}}
\newcommand {\change} {\leftrightarrow}
\newcommand {\ra} {\rightarrow}
\newcommand {\larrow} {\leftarrow}
\newcommand {\ul}   {\underline}
\newcommand {\pr}   {{\quad .}}
\newcommand {\com}  {{\quad ,}}
\newcommand {\q}    {\quad}
\newcommand {\qq}   {\quad\quad}
\newcommand {\qqq}   {\quad\quad\quad}
\newcommand {\qqqq}   {\quad\quad\quad\quad}
\newcommand {\qqqqq}   {\quad\quad\quad\quad\quad}
\newcommand {\qqqqqq}   {\quad\quad\quad\quad\quad\quad}
\newcommand {\qqqqqqq}   {\quad\quad\quad\quad\quad\quad\quad}
\newcommand {\lb}    {\linebreak}
\newcommand {\nl}    {\newline}

\newcommand {\vs}[1]  { \vspace*{#1 cm} }

\newcommand {\MPL}  {Mod.Phys.Lett.}
\newcommand {\NP}   {Nucl.Phys.}
\newcommand {\PL}   {Phys.Lett.}
\newcommand {\PR}   {Phys.Rev.}
\newcommand {\PRL}   {Phys.Rev.Lett.}
\newcommand {\IJMP}  {Int.Jour.Mod.Phys.}
\newcommand {\CMP}  {Commun.Math.Phys.}
\newcommand {\JMP}  {Jour.Math.Phys.}
\newcommand {\AP}   {Ann.of Phys.}
\newcommand {\PTP}  {Prog.Theor.Phys.}
\newcommand {\NC}   {Nuovo Cim.}
\newcommand {\CQG}  {Class.Quantum.Grav.}


\font\smallr=cmr5
\newcommand {\npl}  {{\frac{n\pi}{l}}}
\newcommand {\mpl}  {{\frac{m\pi}{l}}}
\newcommand {\kpl}  {{\frac{k\pi}{l}}}

\def\ocirc#1{#1^{^{{\hbox{\smallr\llap{o}}}}}}
\def\ogamma{\ocirc{\gamma}{}}
\def\oM{{\buildrel {\hbox{\smallr{o}}} \over M}}
\def\osigma{\ocirc{\sigma}{}}

\def\overleftrightarrow#1{\vbox{\ialign{##\crcr
 $\leftrightarrow$\crcr\noalign{\kern-1pt\nointerlineskip}
 $\hfil\displaystyle{#1}\hfil$\crcr}}}
\def\overnab{{\overleftrightarrow\nabslash}}

\def\va{{a}}
\def\vb{{b}}
\def\vc{{c}}
\def\tilpsi{{\tilde\psi}}
\def\tbpsi{{\tilde{\bar\psi}}}

\def\delL{{\delta_{LL}}}
\def\delG{{\delta_{G}}}
\def\delc{{\delta_{cov}}}

\newcommand {\sqxx}  {\sqrt {x^2+1}}   
\newcommand {\gago}  {\gamma^5}
\newcommand {\Pp}  {P_+}
\newcommand {\Pm}  {P_-}
\newcommand {\GfMp}  {G^{5M}_+}
\newcommand {\GfMpm}  {G^{5M'}_-}
\newcommand {\GfMm}  {G^{5M}_-}
\newcommand {\Omp}  {\Omega_+}    
\newcommand {\Omm}  {\Omega_-}
\def\Aslash{{}\hbox{\hskip2pt\vtop
 {\baselineskip23pt\hbox{}\vskip-24pt\hbox{/}}
 \hskip-11.5pt $A$}}
\def\Rslash{{}\hbox{\hskip2pt\vtop
 {\baselineskip23pt\hbox{}\vskip-24pt\hbox{/}}
 \hskip-11.5pt $R$}}

\def\kslash{
{}\hbox       {\hskip2pt\vtop
                   {\baselineskip23pt\hbox{}\vskip-24pt\hbox{/}}
               \hskip-8.5pt $k$}
           }    
\def\qslash{
{}\hbox       {\hskip2pt\vtop
                   {\baselineskip23pt\hbox{}\vskip-24pt\hbox{/}}
               \hskip-8.5pt $q$}
           }    
\def\dslash{
{}\hbox       {\hskip2pt\vtop
                   {\baselineskip23pt\hbox{}\vskip-24pt\hbox{/}}
               \hskip-8.5pt $\partial$}
           }    
\def\dbslash{{}\hbox{\hskip2pt\vtop
 {\baselineskip23pt\hbox{}\vskip-24pt\hbox{$\backslash$}}
 \hskip-11.5pt $\partial$}}

\def\Kbslash{{}\hbox{\hskip2pt\vtop
 {\baselineskip23pt\hbox{}\vskip-24pt\hbox{$\backslash$}}
 \hskip-11.5pt $K$}}
\def\Ktilbslash{{}\hbox{\hskip2pt\vtop
 {\baselineskip23pt\hbox{}\vskip-24pt\hbox{$\backslash$}}
 \hskip-11.5pt ${\tilde K}$}}
\def\Ltilbslash{{}\hbox{\hskip2pt\vtop
 {\baselineskip23pt\hbox{}\vskip-24pt\hbox{$\backslash$}}
 \hskip-11.5pt ${\tilde L}$}}
\def\Qtilbslash{{}\hbox{\hskip2pt\vtop
 {\baselineskip23pt\hbox{}\vskip-24pt\hbox{$\backslash$}}
 \hskip-11.5pt ${\tilde Q}$}}
\def\Rtilbslash{{}\hbox{\hskip2pt\vtop
 {\baselineskip23pt\hbox{}\vskip-24pt\hbox{$\backslash$}}
 \hskip-11.5pt ${\tilde R}$}}
\def\Kbarbslash{{}\hbox{\hskip2pt\vtop
 {\baselineskip23pt\hbox{}\vskip-24pt\hbox{$\backslash$}}
 \hskip-11.5pt ${\bar K}$}}
\def\Lbarbslash{{}\hbox{\hskip2pt\vtop
 {\baselineskip23pt\hbox{}\vskip-24pt\hbox{$\backslash$}}
 \hskip-11.5pt ${\bar L}$}}
\def\Rbarbslash{{}\hbox{\hskip2pt\vtop
 {\baselineskip23pt\hbox{}\vskip-24pt\hbox{$\backslash$}}
 \hskip-11.5pt ${\bar R}$}}
\def\Qbarbslash{{}\hbox{\hskip2pt\vtop
 {\baselineskip23pt\hbox{}\vskip-24pt\hbox{$\backslash$}}
 \hskip-11.5pt ${\bar Q}$}}
\def\Acalbslash{{}\hbox{\hskip2pt\vtop
 {\baselineskip23pt\hbox{}\vskip-24pt\hbox{$\backslash$}}
 \hskip-11.5pt ${\cal A}$}}

\begin{flushright}
May 2004\\
US-04-03\\
hep-th/0405065
\end{flushright}

\vspace{0.5cm}

\begin{center}

{\Large\bf 
Family of Singular Solutions in
A SUSY Bulk-Boundary System
}

\vspace{1.5cm}
{\large Shoichi ICHINOSE
         \footnote{
E-mail address:\ ichinose@u-shizuoka-ken.ac.jp
                  }
}\ and\ 
{\large Akihiro MURAYAMA$^\ddag$
         \footnote{
E-mail address:\ edamura@ipc.shizuoka.ac.jp
                  }
}
\vspace{1cm}

{\large 
Laboratory of Physics, 
School of Food and Nutritional Sciences, 
University of Shizuoka, 
Yada 52-1, Shizuoka 422-8526, Japan
 }

$\mbox{}^\ddag${\large
Department of Physics, Faculty of Education, Shizuoka University,
Shizuoka 422-8529, Japan
}
\end{center}

\vfill

{\large Abstract}\nl
A set of classical solutions of a singular type
is found in a 5D SUSY bulk-boundary system.
The "parallel" configuration, where the whole components
of fields or branes are parallel in the iso-space, 
naturally appears. 
It has three {\it free} parameters related to
the {\it scale freedom} in the choice of the 
brane-matter sources and the {\it "free" wave} property
of the {\it extra component} of the bulk-vector field.
The solutions describe brane, anti-brane and brane-anti-brane
configurations depending on the parameter choice.
Some solutions describe the localization behaviour
even after the non-compact limit of the extra space. 
Stableness is assured. 
Their meaning in the brane world physics is
examined in relation to the stableness, localization, non-singular
(kink) solution
and the bulk Higgs mechanism.

\vspace{0.5cm}

PACS NO: 
\ 11.10.Kk,
\ 11.27.+d,
\ 12.60.Jv,
\ 12.10.-g,
\ 11.25.Mj 
\nl

\vspace{3mm}
{\bf 1}\ {\it Introduction}\q
In the soliton physics, the kink solution is the simplest
example to show the characteristic properties of the soliton:\ 
energy localization, stability, asymptotic vacua, 
conserved quantity (index), etc..(See, for example, a nice textbook
by R. Rajaraman\cite{Rajara82}.)
\begin{eqnarray}
\phi_{kink}(y)=\phi_0\tanh (ky)\com\q  -\infty <y<\infty
\com\label{int1}
\end{eqnarray}
where $\phi_0$ and $k$ are constants. $1/k$ corresponds to
the "{\it thickness}" parameter in the brane world. 
This solution is $Z_2$-odd:\ 
$\phi_{kink}(y)=-\phi_{kink}(-y)$. It is a {\it stable} vacuum
solution of the 1+1 dim scalar field theory
with the Higgs potential. 
\begin{eqnarray}
\Lcal=-\half\pl_\m\phi\pl^\m\phi
-\frac{\la}{4}(\phi^2-\vz^2)^2\com\q
\phi_0=\vz>0\com\q\la>0\com\q k=\sqrt{\frac{\la}{2}}\,\vz\com
\nn
(x^\m)=(x^0=t, x^1=y)\com\q (\eta^\mn )=(-,+)
\pr\label{int2}
\end{eqnarray}
This is a typical model of the spontaneous symmetry breaking.
The symmetry, in this simple example, is 
$Z_2$-symmetry (a discrete symmetry):\ 
$y\change -y$. 
The {\it stableness} is guaranteed by that the kink solution
connects two degenerate vacua:\ $\phi=\phi_0$ at $y=\infty$ and 
$\phi=-\phi_0$ at $y=-\infty$. 
For the static configuration on this background, 
the leading value of the action, $S$, 
in the "{\it thin-wall}" limit
($kL\gg 1$, $L$: infrared regularization parameter of y-axis)
:\ 
$\phi_{kink}\sim\phi_0{\tilde \ep}(ky),\ 
\pl_y\phi_{kink}\sim 2\phi_0{\tilde \del}(y)$
,is estimated as, 
\begin{eqnarray}
\vp\equiv \phi-\phi_{kink}\com\q |\vp|\ll 1\com\nn
\Lcal\approx -\half \pl_y(\phi_{kink}+\vp)\pl_y(\phi_{kink}+\vp)
\approx -2 {\phi_0}^2{\tilde \del}(y)^2
-2\phi_0{\tilde \del}(y)\pl_y\vp+O(\vp^2)\com\nn
S=\int dy dt\Lcal\approx -2\int dt \phi_0^2{\tilde \del}(0)
=-\frac{1}{\pi}L\cdot T\phi_0^2\com\nn
S/LT\approx -\frac{1}{\pi}\phi_0^2
\com\label{int3}
\end{eqnarray}
where the {\it infrared regularization} 
parameters, $L$ and $T$, are introduced:\ 
$-T/2<t<T/2,\, -L/2<y<L/2$. ${\tilde \ep}(y)$ and ${\tilde \del}(y)$
are the ordinary (non-periodic) sign and delta functions respectively.
\footnote{
${\tilde \ep}(y)$ and ${\tilde \del}(y)$ are defined by
\begin{eqnarray}
{\tilde \ep(y)}=\left\{
\begin{array}{cc}
+1 & \mbox{for }y>0 \\
0  & \mbox{for }y=0 \\
-1 & \mbox{for }y<0 \end{array}
\right.\q\q \com\q\q
\int_{-\infty}^{\infty}{\tilde \del}(y)f(y)dy=f(0)
\pr
\label{int3b}
\end{eqnarray}
These should be compared with periodic ones, $\ep(y)$ and $\del(y)$, 
used later.
}

In the recent development of the brane world,
it has become clear that the kink-type configuration
plays a very important role in the {\it extra-space} behaviour
of the higher dimensional models. 
This is because it describes 
the stable localization configuration.
In the Randall-Sundrum model
I(wall-anti-wall model)\cite{RS9905}, they considered the following
{\it bulk-boundary} theory in the AdS$_5$ space-time on 
$S^1/Z_2$ orbifold. 
\begin{eqnarray}
S=\int d^4x\int_{-\pi}^{\pi}d\xf
\sqrt{-G}\{ 
-\La+2M^3R-\del(\xf)V_{hid}-\del(\xf-\pi)V_{vis}
          \}        \com\nn
-\pi<\xf\leq \pi\com\q
          ds^2=\e^{-2\si(\xf)}\eta_\mn dx^\m dx^\n
          +\frac{l^2}{\pi^2}(d\xf)^2
\com\label{int4}
\end{eqnarray}
where $\La,M,V_{hid(vis)}$ are 5D cosmological constant,
5D Planck mass and the brane tension at $\xf=0(\pi)$. 
$\del(s)$ is the {\it periodic} delta function. 
The Einstein equation 
and $Z_2$-symmetry (even) of $\si(\xf)$
requires
\begin{eqnarray}
\si=\frac{l}{\pi}\sqrt{\frac{-\La}{24M^3}}\,|\xf|\com\q \La<0\com\q
\si''=\frac{2l}{\pi}\sqrt{\frac{-\La}{24M^3}}
(\del(\xf)-\del(\xf-\pi))\com\nn
V_{hid}=-V_{vis}=24M^3k\com\q \La=-24M^3k^2\com
\label{int5}
\end{eqnarray}
where $k$ is a scale with mass dimension. 
They applied this result to the mass hierarchy problem
and give rich possibilities in the unified models.
In the Randall-Sundrum model II\cite{RS9906} (one wall model), 
partly from
the stability assurance, they considered the $l\ra \infty$
limit of the model I.
\begin{eqnarray}
S=\int d^4x\int_{-\infty}^{\infty}d\xf
\sqrt{-G}\{ 
-\La+2M^3R-{\tilde \del}(\xf)V
          \}        \com\nn
          ds^2=\e^{-2\si(\xf)}\eta_\mn dx^\m dx^\n
          +(d\xf)^2
   \com\q -\infty<x^\m,\ x^5<\infty\com\nn
\si=\sqrt{\frac{-\La}{24M^3}}\,|\xf|\com\q \La<0\com\q
\si''=2\sqrt{\frac{-\La}{24M^3}}
\del(\xf)
\pr\label{int6}
\end{eqnarray}
In this model, the {\it stability} is guaranteed by
the same reason as the first example of the kink solution.
\footnote{
Another way out was
 suggeted in \cite{RS9905} and was analysed
 by Goldberger and Wise\cite{GW9907}. They 
try to stabilize the system, keeping the compact extra-space, 
by regarding the length parameter
$l$ as an expectation value of some scalar field (radion).
}
In fact the above solution can be obtained by
the "thin-wall" limit of the {\it generalized} kink solution
in the {\it bulk} Higgs model \cite{SI0003,SI0107}.
\begin{eqnarray}
S=\int d^5X\sqrt{-G}(-\half M^3R
-\half G^{AB}\pl_A\Phi\pl_B\Phi-V(\Phi))\ ,\ 
V(\Phi)=\frac{\la}{4}(\Phi^2-\vz^2)^2+\La
\ .\label{int7}
\end{eqnarray}
This model makes it possible to treat 
the brane system in the non-singular way.

Both models explained above are non-supersymmetric. The first one
is a flat theory, whereas the second one is the model of
a 5D gravitational space-time (AdS$_5$). 
The "curvedness" simply comes from the "warped" factor, $\e^{-2\si}$. 
For a fixed $\xf$-slice, the 4D space-time is the flat one. 
The {\it dilaton} ($\si$)
field part controls the bulk-scale of the each 4D Minkowski
(flat) slice at the point $\xf$. In the present paper, 
we examine a 5D SUSY {\it flat} theory where {\it two} scalar fields
, which come from the 5D SUSY multiplet, 
play the similar role to the scalar and dilaton fields in the
above two examples.

{\bf 2}\ {\it Mirabelli-Peskin Model}\q
Inspired by the Horava-Witten model\cite{HW96} (11D supergravity
on $S_1/Z_2$-orbifold, the strong coupling limit of
the 10D heterotic string theory), Mirabelli and Peskin\cite{MP97}
proposed a field theory model describing a bulk-boundary
system which mimics the brane(-anti-brane) configuration
in the string theory. 
Let us consider the 5 dimensional flat space-time with the signature
(-1,1,1,1,1).
\footnote{
Notation is basically the same as ref.\cite{Hebec01}.
} 
The space of the fifth
component is taken to be $S_1$, 
with the periodicity $2l$, and has the $Z_2$-orbifold condition.
\begin{eqnarray}
\xf\ra\xf+2l\ (\mbox{periodicity})\com\q
\xf\change -\xf\ (Z_2\mbox{-symmetry})\pr
\label{mp1b}
\end{eqnarray}
We take a 
5D bulk theory $\Lcal_{bulk}$ which is
coupled with a 4D matter theory $\Lcal_{bnd}$ on a "wall" at $\xf=0$
and with $\Lcal'_{bnd}$ on the other "wall" at $\xf=l$.
\begin{eqnarray}
S=\int_{-l}^{l}dx^5\int d^4x\{\Lcal_{blk}+\del(x^5)\Lcal_{bnd}+\del(x^5-l){\Lcal'}_{bnd}
\}
\pr\label{mp1}
\end{eqnarray}

The bulk dynamics is given by the 5D super YM theory
which is made of 
a vector field $A^M\ (M=0,1,2,3,5)$, 
a scalar field $\Phi$, 
a doublet of symplectic Majorana fields $\la^i\ (i=1,2)$, 
and a triplet of auxiliary scalar fields $X^a\ (a=1,2,3)$:
\begin{eqnarray}
\Lcal_{SYM}=-\half\tr {F_{MN}}^2-\tr (\na_M\Phi)^2
-i\tr(\labar_i\ga^M\na_M\la^i)
+\tr (X^a)^2+g\,\tr (\labar_i[\Phi,\la^i])\com   
\label{mp2}
\end{eqnarray}
where all bulk fields are the {\it adjoint} representation
(suffixes: $\al,\be,\cdots $)
of the gauge group $G$. 
The bulk Lagrangian $\Lcal_{SYM}$ 
is invariant under the 5D SUSY transformation.
This system has the symmetry of
8 real super charges.

It is known that we can consistently project out $\Ncal=1$ SUSY
multiplet, which has 4 real super charges, 
by assigning $Z_2$-parity 
to all fields in accordance with the 5D SUSY. 
A consistent choice is given as:\  $P=+1$ for 
$A^m, \la_L, X^3$;  $P=-1$ for 
$A^5, \Phi, \la_R, X^1, X^2$ ($m=0,1,2,3$). 
Then ($A^m,\la_L,X^3-\na_5\Phi$) constitute
 an $\Ncal =1$ vector multiplet. 
Especially $\Dcal\equiv X^3-\na_5\Phi$ plays the role
of {\it D-field} on the wall. 
We introduce one 4D chiral multiplet ($\phi,\psi,F$) on the $\xf=0$ wall
and the other one ($\phi',\psi',F'$) on the $\xf=l$ wall: 
complex scalar fields $\phi,\phi'$, Weyl spinors $\psi,\psi'$, and
auxiliary fields of complex scalar $F,F'$. 
These are the simplest matter candidates and were taken
in the original theory\cite{MP97}. 
Using the $\Ncal=1$ SUSY property of the fields 
($A^m,\la_L,X^3-\na_5\Phi$),
we can find the following bulk-boundary coupling on the $\xf=0$ wall.
\begin{eqnarray}
\Lcal_{bnd}=-\na_m\phi^\dag \na^m\phi-\psi^\dag i\sibar^m \na_m\psi+F^\dag F\nn
+\sqrt{2}ig(\psibar\labar_L\phi-\phi^\dag\la_L\psi)
+g\phi^\dag \Dcal\phi
+\Lcal_{SupPot}
\ ,\nn
\Lcal_{SupPot}=
(\half m_{\alp\bep}\Th_\alp\Th_\bep
+\frac{1}{3!}\la_{\alp\bep\gap}\Th_\alp\Th_\bep\Th_\gap)|_{\theta^2}
+\mbox{h.c.}
\com
\label{mp7}
\end{eqnarray}
where $    
\na_m\equiv \pl_m+igA_m,\ \Dcal =X^3-\na_5\Phi,\ 
\Th=\phi+\sqtwo\theta\psi+\theta^2 F
$.
We take the {\it fundamental} representation for $\Th=(\phi,\psi,F)$. 
The quadratic (kinetic) terms of the vector $A^m$, the gaugino spinor $\la_L$
and the 'auxiliary' field $\Dcal=X^3-\na_5\Phi$ are in the bulk world. 
In the same way we introduce the coupling between the matter fields
($\phi',\psi',F'$) on the $\xf=l$ wall and the bulk fields:\ 
$\Lcal'_{bnd}=(\phi\ra\phi', \psi\ra\psi', F\ra F' \ in\ \mbox{(\ref{mp7})})$. 
We note the interaction between the bulk fields and the boundary
ones is {\it definitely fixed from SUSY}.

{\bf 3}\ {\it Vacuum of Mirabelli-Peskin Model}\q
We now examine the vacuum structure. 
Generally the vacuum is
determined by the potential part of scalar fields.
We first reduce the previous system to the part
which involves only scalar fields or the extra
component of the bulk vector. 
\begin{eqnarray}
\Lcal^{red}_{blk}[\Phi,X^3,A_5]
=\tr \left\{ -\pl_M\Phi\pl^M\Phi+X^3X^3-\pl_MA_5\pl^MA_5
+2g(\pl_5\Phi\times A_5)\Phi\right.\nn
\left. -g^2(A_5\times\Phi)(A_5\times\Phi)
-2\pl_M\cbar\cdot\pl^Mc-2ig\pl_5\cbar\cdot [A^5,c]\right\}
+\mbox{irrel. terms}
\com
\label{ep3}
\end{eqnarray}
where we have dropped terms of $2\tr X^1X^1=X^1_\al X^1_\al, 
2\tr X^2X^2=
X^2_\al X^2_\al$
as 'irrelevant terms' because they decouple from other fields. 
(Note $\tr (\pl_5\Phi\times A_5)\Phi
=(1/2)f_{\ab\ga}\pl_5\Phi_\al A_{5\be}\Phi_\ga$). 
The field $c$ is the ghost field which is introduced in the
usual procedure of fixing the gauge freedom of $\Lcal_{SYM}$.
While
$\Lcal_{bnd}$, on the $\xf=0$ wall, reduces to
\begin{eqnarray}
\Lcal^{red}_{bnd}[\phi,\phi^\dag,X^3-\na_5\Phi]=-\pl_m\phi^\dag\pl^m\phi
+g(X^3_\al-\na_5\Phi_\al)\phi^\dag_\bep (T^\al)_{\bep\gap}\phi_\gap
+F^\dag F
\nn
+
\left\{
\frac{m_{\alp\bep}}{2}(\phi_\alp F_\bep+F_\alp\phi_\bep)
+\frac{\la_{\alp\bep\gap}}{3!}(\phi_\alp\phi_\bep F_\gap
+\phi_\alp F_\bep\phi_\gap+F_\alp\phi_\bep\phi_\gap)+\mbox{h.c.}
\right\}
\ .
\label{ep4}
\end{eqnarray}
$\alp, \bep\cdots$ are the suffixes of the fundamental representation.
In the same way, we obtain 
${\Lcal^{red}_{bnd}}'[\phi',{\phi'}^\dag,X^3-\na_5\Phi]$
on the $\xf=l$ wall by replacing, in (\ref{ep4}), 
$\phi$ and $\phi^\dag$ by $\phi'$ and $\phi'^\dag$, respectively.

The vacuum is usually obtained by the
{\it constant} solution of the scalar-part field equation. 
In higher dimensional models, however, 
extra-coordinate(s) can be regarded as parameter(s)
which should be separately treated 
from the 4D space-time coordinates. 
In this standpoint, it is the more general treatment of
the vacuum that we allow the {\it $\xf$-dependence} on 
the bulk-part of the solution. 
We generally call the classical solutions
($\vp, \chi^3, a_5; \eta, \eta', f, f'$) 
the {\it background fields}. 
\footnote{
In the background field treatment\cite{IM0312} we 
expand all fields around the background fields:\ 
$\vp+\Phi, \chi^3+X^3, a_5+A_5, \eta+\phi, \eta'+\phi', 
f+F, f'+F'$
}
They satisfy 
the {\it field equations} derived from (\ref{ep3}) and (\ref{ep4})\ 
({\it on-shell} condition). 
Assuming 
$\vp=\vp(\xf), \chi^3=\chi^3(\xf), 
a_5=a_5(\xf), \eta=\mbox{const}, \eta'=\mbox{const},
f=\mbox{const}, f'=\mbox{const}$, 
the field equation of 
$
\Lcal^{red}_{blk}+\del(x^5)\Lcal^{red}_{bnd}
+\del(x^5-l){\Lcal^{red}_{bnd}}'
$
are given by, for the bulk-fields variation, 
\begin{eqnarray}
\del\Phi_\al\q;\nn
{\pl_5}^2\vp_\al+gf_{\be\ga\al}\pl_5\vp_\be a_{5\ga}
-gf_{\ab\ga}\pl_5(a_{5\be}\vp_\ga)
-g^2f_{\be\al\tau}f_{\ga\del\tau}a_{5\be}a_{5\ga}\vp_\del
\nn
+g\pl_5(\del(\xf))\eta^\dag T^\al\eta
+g\pl_5(\del(\xf-l))\eta'^\dag T^\al\eta'
+g^2(\del(\xf)\eta^\dag T^\ga\eta+\del(\xf-l)\eta'^\dag T^\ga\eta')
f^{\be\al\ga}a_{5\be}\nn
=-\pl_5Z_\al-g(Z\times a_5)_\al=0\ ,\label{vac1a}
\\
\nn
\del A_{5\al}\q;\nn
{\pl_5}^2a_{5\al}+gf_{\be\al\ga}\pl_5\vp_\be\, \vp_\ga
-g^2f_{\ab\tau}f_{\ga\del\tau}\vp_\be a_{5\ga}\vp_\del
+g^2(\del(\xf)\eta^\dag T^\ga\eta+\del(\xf-l)\eta'^\dag T^\ga\eta')
f^{\al\be\ga}\vp_\be\nn
={\pl_5}^2a_{5\al}-g(\vp\times Z)_\al=0\ ,\label{vac1b}
\\
\nn
\del X^3_\al\q;\nn
\chi^3_\al+g(\del(\xf)\eta^\dag T^\al \eta+\del(\xf-l)\eta'^\dag T^\al \eta')
=0\ ,
\com
\label{vac1c}
\end{eqnarray}
where 
$Z_\al\equiv -g(\del(\xf)\eta^\dag T^\al\eta+\del(\xf-l)\eta'^\dag T^\al\eta')
-\pl_5\vp_\al+gf_{\ab\ga}a_{5\be}\vp_\ga$. 
The field equations for 
the boundary-fields part are given by
\begin{eqnarray}
\del\phi^\dag_\alp\q (\del\phi'^\dag_\alp)\q;\nn
{d_\be}|_{\xf=0}\times (T^\be\eta)_\alp
+m_{\alp\bep}f^\dag_\bep+\half\la_{\alp\bep\gap}\eta^\dag_\bep f^\dag_\gap
=0\ ,\ 
(\eta\ra\eta', f\ra f'\ \mbox{in the left equation})\ ,\ \label{vac2a}
\\
\nn
\del F^\dag_\alp\q (\del F'^\dag_\alp)\nn
f_\alp+m_{\alp\bep}\eta^\dag_\bep+\half\la_{\alp\bep\gap}
\eta^\dag_\bep\eta^\dag_\gap=0\ ,\ 
(\eta\ra\eta', f\ra f'\ \mbox{in the left equation})
\com\ 
\label{vac2b}
\end{eqnarray}
where $d_\al=(\chi^3-\pl_5\vp+ga_5\times\vp)_\al$ 
is the background D-field. 
From the equation (\ref{vac1c}), we obtain
\begin{eqnarray}
\chi^3_\al=-g(\del(\xf)\eta^\dag T^\al \eta+\del(\xf-l)\eta'^\dag T^\al \eta')
\pr
\label{vac3}
\end{eqnarray}
Then we know
\begin{eqnarray}
Z_\al=d_\al
\pr
\label{vac4}
\end{eqnarray}

Before systematically solving the equations above,
we note a simple structure involved in them. Under
the "{\it parallel}" circumstance, 
$a_{5\al}\propto\vp_\al\propto\eta^\dag T^\al\eta
\propto\eta'^\dag T^\al\eta'$, the equations
for $\del\Phi_\al$ (\ref{vac1a})
and $\del A_{5\al}$ (\ref{vac1b})
are
\begin{eqnarray}
{\pl_5}^2\vp_\al=
-g\pl_5(\del(\xf)\eta^\dag T^\al\eta
+\del(\xf-l)\eta'^\dag T^\al\eta')\com\nn
{\pl_5}^2a_{5\al}=0
\pr
\label{vac4b}
\end{eqnarray}
The first one is a static wave equation with "source" fields
located at $\xf=0$ and $l$. 
It is easily integrated once. 
\begin{eqnarray}
{\pl_5}\vp_\al=
-g(\del(\xf)\eta^\dag T^\al\eta
+\del(\xf-l)\eta'^\dag T^\al\eta')+\mbox{const}
\pr
\label{vac4c}
\end{eqnarray}
This result was used in the original paper\cite{MP97}. 
The second equation of (\ref{vac4b}) is  a (static)
"free" wave equation (no source fields). $a_{5\al}$
do {\it not} receive, in the "parallel" environment, any
effect from the boundary sources $\eta, \eta'$.
This characteristically shows the difference between
the bulk scalar $\Phi_\al$ and the extra component
of the bulk vector $A_{5\al}$ in the vacuum
configuration. 

We first solve (\ref{vac1a}),(\ref{vac1b}) and (\ref{vac1c}) with respect to
$a_{5\al}$ and $\vp_\al$. They also give the solutions
for $\chi^3_\al$ and $d_\al=Z_\al$. Using these results
we solve (\ref{vac2a}) and (\ref{vac2b}) with respect to
$\eta, \eta', f$ and $f'$ for given values of
$m_{\alp\bep}$ and $\la_{\alp\bep\gap}$. 
Here we seek a natural solution by requiring that 
{\it $d_\al$ is independent of $\xf$}. 
\begin{eqnarray}
Z_\al=d_\al=-g(\del(\xf)\eta^\dag T^\al\eta+\del(\xf-l)\eta'^\dag T^\al\eta')
-\pl_5\vp_\al+gf_{\ab\ga}a_{5\be}\vp_\ga=\nn
\mbox{independent of }\xf\mbox{ (const)}
\pr
\label{vac5}
\end{eqnarray}
Then, from the equation of (\ref{vac1a}), we have 
$Z\times a_5=0$.  It  says that 
we may consider the three cases :\ 
1)\ $a_{5\al}=0$, \ 2)\ $Z_\al=0$, \ 3)\ $a_{5\al}\propto Z_\al (\neq 0)$. 
It turns out that the case 3) includes the case 1) and 2). Hence we explain
case 3). 

Before proceeding the analysis furthermore, we note here
a mathematical fact about the solution of the "free" field equation
in $S^1/Z_2$ space.
\begin{eqnarray}
\frac{d^2}{d y^2}f(y)=0\ \mbox{except the fixed point}(y=0)
\mbox{ and the periodic point****}(y=l)\com\nn
\mbox{periodicity}:\ f(y)=f(y+2l)\com\nn 
\mbox{A) }Z_2-\mbox{odd}:\ f(y)=-f(-y)\com\q
\mbox{B) }Z_2-\mbox{even}:\ f(y)=f(-y)
\pr
\label{vac8}
\end{eqnarray}
\nl
A)\ $Z_2$-odd\nl
The two independent solutions are given by
the periodic sign function (see Fig.1)
\begin{eqnarray}
\ep(y)=\left\{
\begin{array}{cc}
+1 & \mbox{for }2nl<y<(1+2n)l \\
0  & \mbox{for }y=nl \\
-1 & \mbox{for }(2n-1)l<y<2nl \end{array}
\right.\q n\in {\bf Z}\pr
\label{vac9}
\end{eqnarray}
and the sawtooth-wave function (see Fig.2),
\begin{eqnarray}
 \mbox{[$y$]$_p$} =
\left\{
\begin{array}{cc}
y & -l<y <l \\
0   &   y=l \\
\mbox{periodic} & \mbox{other regions} 
\end{array}
\right.\pr
\label{vac10}
\end{eqnarray}
Both functions are piece-wise continuous. 
A useful relation is 
$[y-l]_p=[y]_p-l\ep(y)$. 
Their derivatives are given by
\begin{eqnarray}
B\Bbar-\mbox{type}:\q
\frac{d\ep(y)}{dy}=2(\del(y)-\del(y-l))\com\nn
\Bbar-\mbox{type}:\q
\frac{d}{dy}[y]_p=1-2l\del(y-l)\com\nn
B-\mbox{type}:\q
-\frac{d}{dy}[y-l]_p=
-\frac{d}{dy}\{ [y]_p-l\ep(y)\}=-1+2l\del(y)
\com
\label{vac11}
\end{eqnarray}
where $\del(y)$ is the periodic (periodicity $2l$) delta function.
We have named the above three distributions Brane-Anti-Brane($B\Bbar$),
Anti-Brane($\Bbar$) and Brane($B$) respectively for a later purpose.
See Fig.3. 
\begin{figure}
\centerline{ \psfig{figure=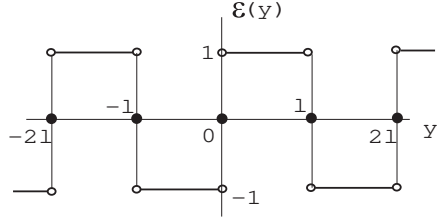,height=3cm,angle=0}}
\caption{ 
The graph of the periodic sign function $\ep(y)$, (\ref{vac9}).
}
\label{fig:Sign2}
\end{figure}
\begin{figure}
\centerline{ \psfig{figure=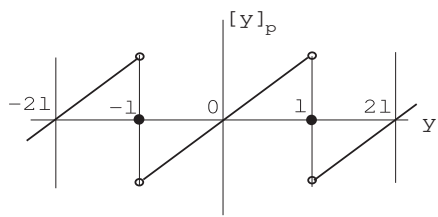,height=3cm,angle=0}}
\caption{ 
The graph of the sawtooth wave $[y]_p$, (\ref{vac10}).
}
\label{fig:SawTooth2}
\end{figure}
\begin{figure}
\centerline{ \psfig{figure=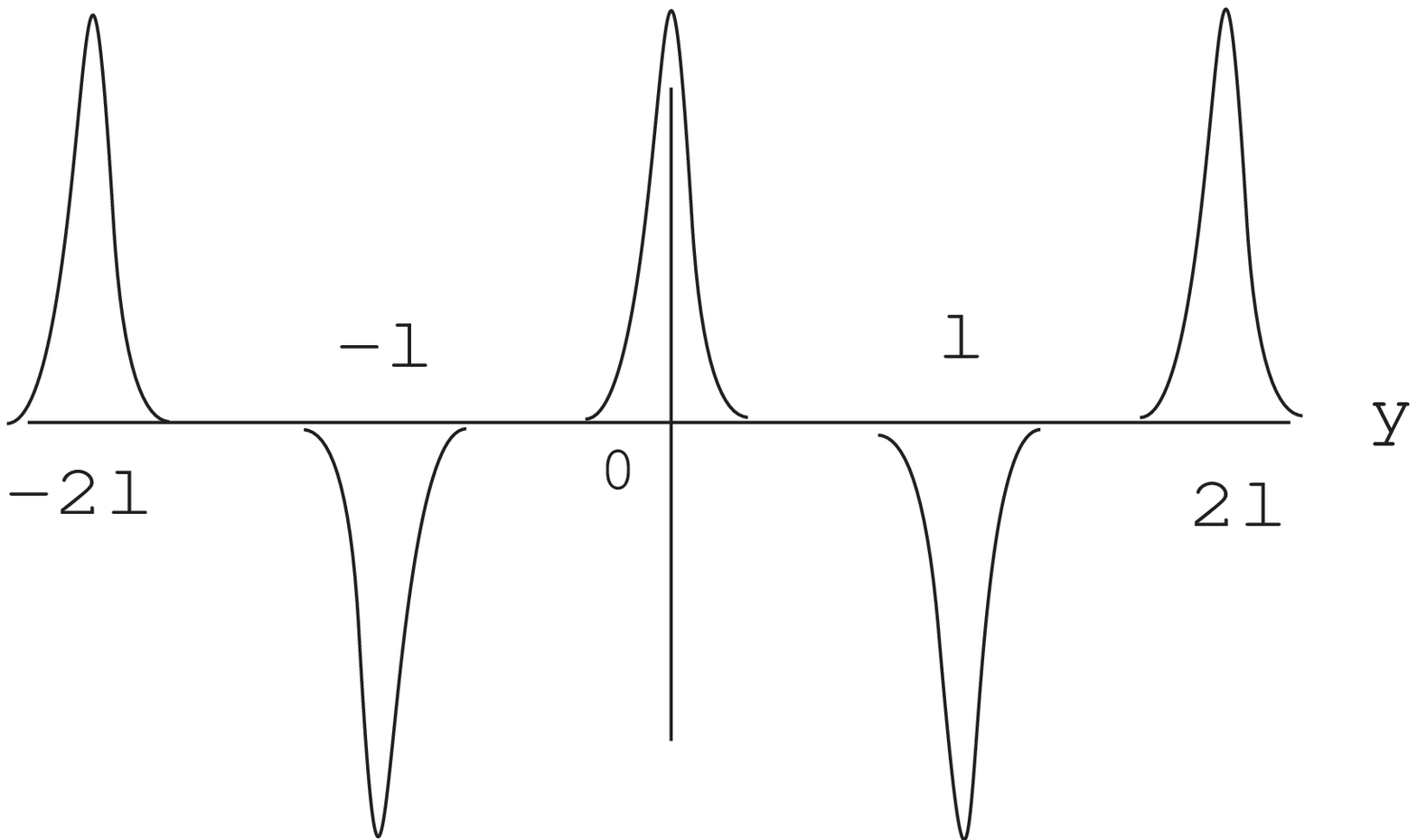,height=3cm,angle=0}}
\centerline{ \psfig{figure=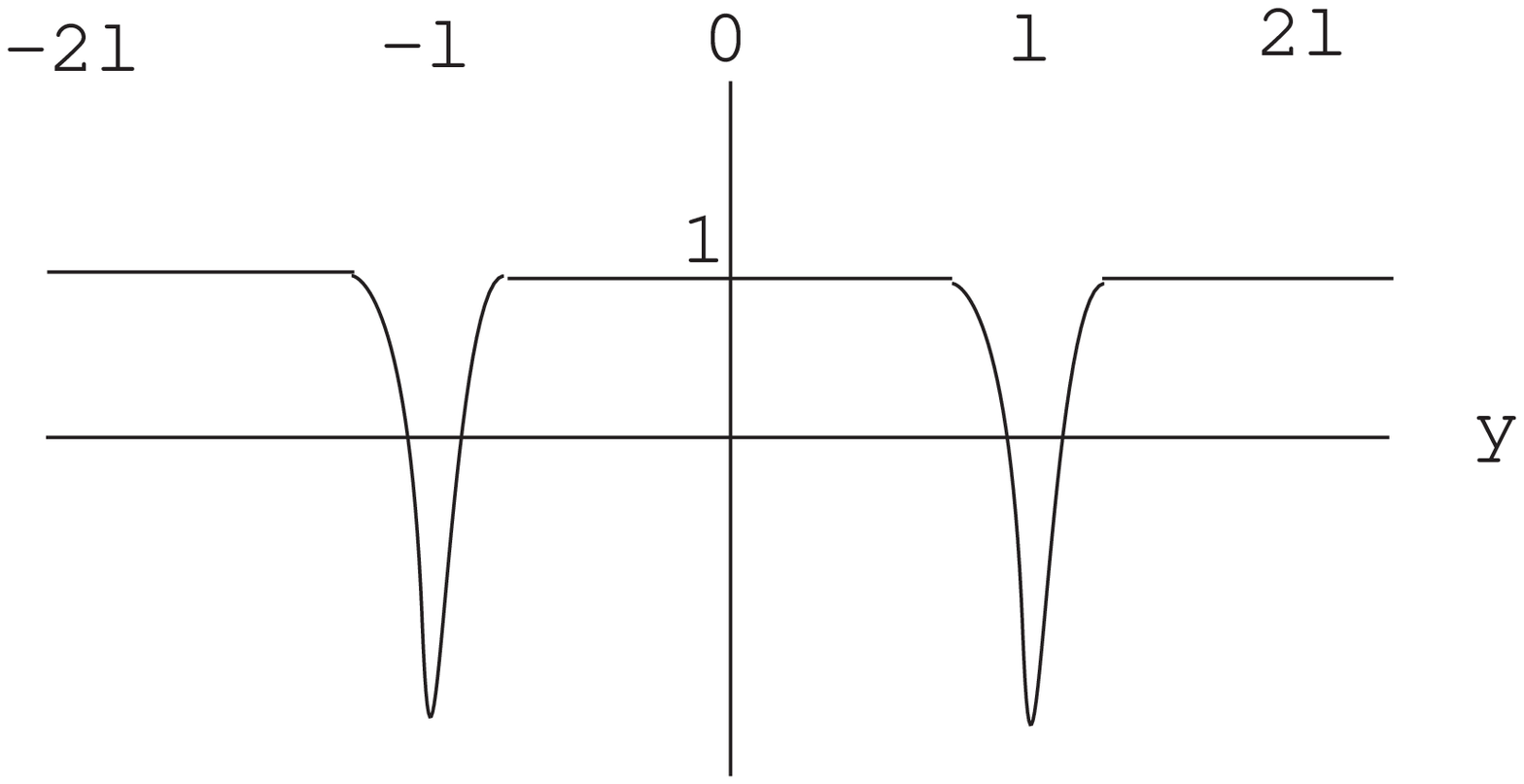,height=3cm,angle=0}}
\centerline{ \psfig{figure=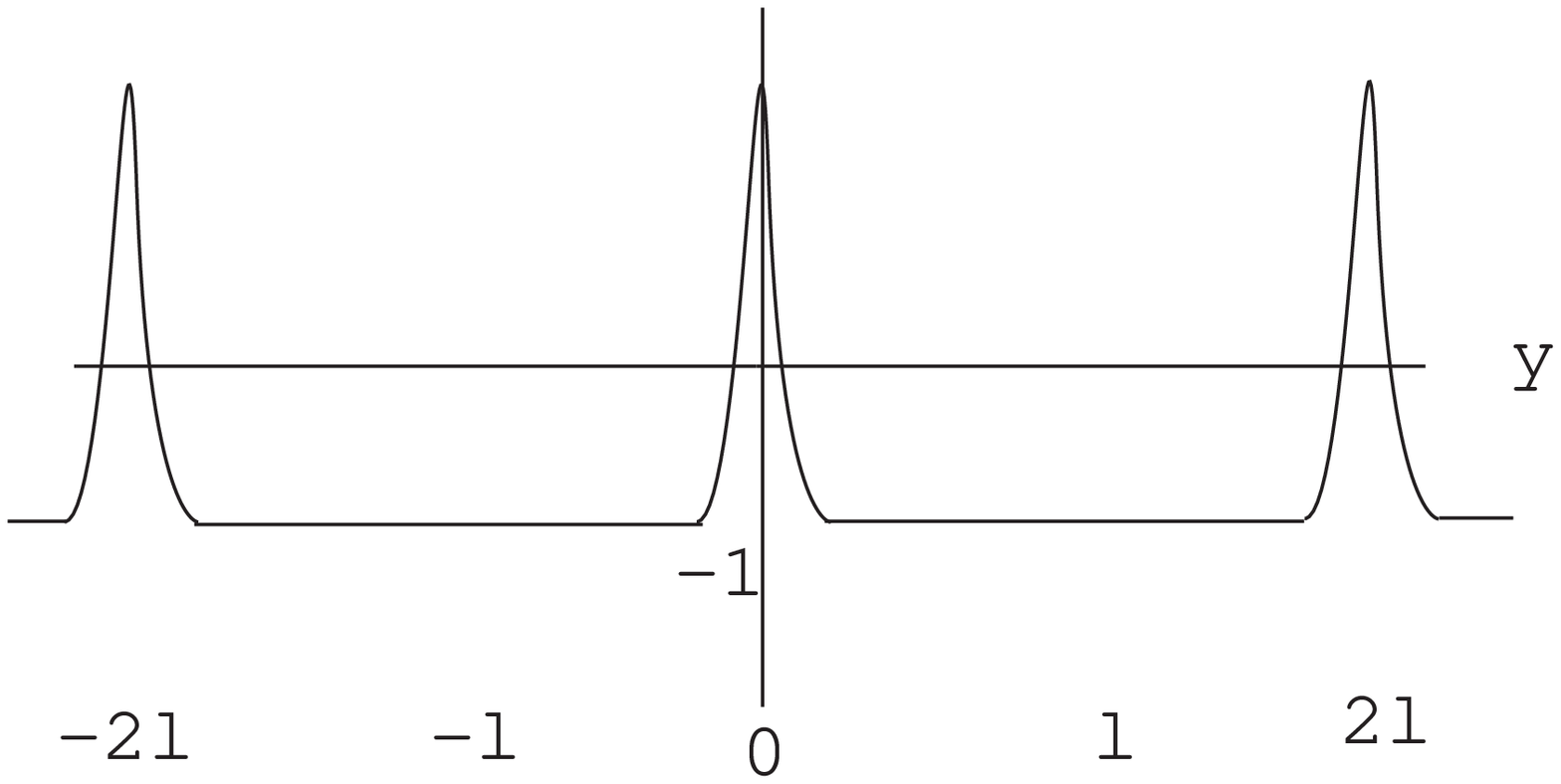,height=3cm,angle=0}}
\caption{ 
The graphs of the three distributions of (\ref{vac11}).
Top:\ Brane-Anti-Brane($B\Bbar$);\ Middle:\ Anti-Brane($\Bbar$); Bottom:
\  Brane($B$).
}
\label{fig:branes}
\end{figure}
\nl
\nl
B)\ $Z_2$-even\nl
The two independent solutions are given by the identity
function (see Fig.4), 
\begin{eqnarray}
 i(y) =
\left\{
\begin{array}{cc}
1 & -l<y \leq l \\
\mbox{periodic} & \mbox{other regions} 
\end{array}
\right.\pr
\label{vac11b}
\end{eqnarray}
and the periodic "absolute-linear" function (see Fig.5), 
\begin{eqnarray}
 v(y) =
\left\{
\begin{array}{cc}
|y| & -l<y \leq l \\
\mbox{periodic} & \mbox{other regions} 
\end{array}
\right.\pr
\label{vac11c}
\end{eqnarray}
Both functions are continuous. 
The first one is smooth and the second one
is piece-wise smooth. 
Their derivatives are given by
\begin{eqnarray}
\frac{di(y)}{dy}=0\com\q
\frac{dv(y)}{dy}=\ep(y)
\pr
\label{vac11d}
\end{eqnarray}
The {\it even} solution $v(y)$ appears as the dilaton in the Randall-Sundrum 
of Sec.1. In the first example of Sec.1 and 
in the present model, the {\it odd} ones appear.
The mathematical fact explained above shows
the important connection among 
the brane configuration, the boundary condition and $Z_2$-symmetry. 
\begin{figure}
\centerline{ \psfig{figure=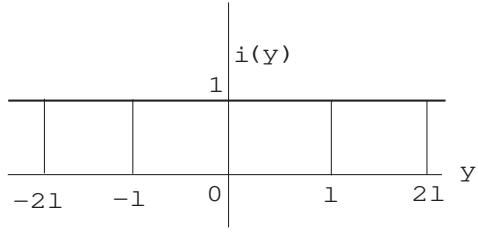,height=3cm,angle=0}}
\caption{ 
The graph of the identity function $i(y)$, (\ref{vac11b}).
}
\label{fig:ident}
\end{figure}
\begin{figure}
\centerline{ \psfig{figure=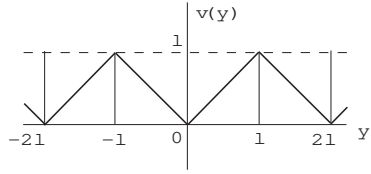,height=3cm,angle=0}}
\caption{ 
The graph of the periodic absolute-linear function $v(y)$, (\ref{vac11c}).
}
\label{fig:PerAbs}
\end{figure}

Let us examine the case 3)\ $a_{5\al}\propto Z_\al (\neq 0)$. 
Noting (\ref{vac5}), we may put the following forms
for $Z_\al$ and $a_{5\al}$. 
\begin{eqnarray}
Z_\al=\Zbar_\al\com\q
a_{5\al}=\abar_\al j(\xf)\com\q
\Zbar_\al\propto \abar_\al(\neq 0)
\com
\label{vac22}
\end{eqnarray}
where $\Zbar_\al$ and $\abar_\al$ are constants and $j(\xf)$ is a function
of $\xf$ which is to be specified below. 
The first equation of (\ref{vac22}) says
\begin{eqnarray}
-g(\del(\xf)\eta^\dag T^\al\eta+\del(\xf-l)\eta'^\dag T^\al\eta')
-\pl_5\vp_\al+g\,j(\xf)(\abar\times \vp)_\al=\Zbar_\al(\mbox{const})
\pr
\label{vac23}
\end{eqnarray}
The equation (\ref{vac1b}) says
\begin{eqnarray}
\abar_\al{\pl_5}^2j-g(\vp\times\Zbar)_\al=0
\pr
\label{vac24}
\end{eqnarray}
First we solve (\ref{vac23}) with the requirement:
\begin{eqnarray}
\vp_\al=\vpbar_\al h(\xf)\com\q
\abar_\al\propto\vpbar_\al
\com
\label{vac25}
\end{eqnarray}
where $\vpbar_\al$ is a constant and $h(\xf)$ is a function of $\xf$
to be determined. 
The second relation says the two scalars, $a_5$ and $\vp$, 
are (anti)parallel in the iso-space. 
Then (\ref{vac23}) reduces to
\begin{eqnarray}
-g(\del(\xf)\eta^\dag T^\al\eta+\del(\xf-l)\eta'^\dag T^\al\eta')
-\vpbar_\al\pl_5h=\Zbar_\al(\mbox{const})
\pr
\label{vac26}
\end{eqnarray}
From the equation (\ref{vac1a}), $h$ satisfies the "free" field equation
except the fixed points. Hence 
we have
\begin{eqnarray}
h=c_1[\xf]_p+[\xf-l]_p \com\nn
\vpbar_\al=\frac{g}{2l}\eta^\dag T^\al\eta=
\frac{1}{c_1}\frac{g}{2l}\eta'^\dag T^\al\eta'\com\nn
Z_\al=d_\al=-\vpbar_\al (1+c_1)
=-\frac{g}{2l}(1+c_1)\eta^\dag T^\al\eta
\pr
\label{vac27}
\end{eqnarray}
where $c_1$ is a free parameter. 
Next we solve (\ref{vac24}). Because $\vp_\al=\vpbar_\al h(\xf)\propto \Zbar_\al$, the equation reduces to the "free" one:\ 
\begin{eqnarray}
\abar_\al{\pl_5}^2j=0\com\q \abar_\al\neq 0
\pr
\label{vac28}
\end{eqnarray}
The solution is 
given by
\begin{eqnarray}
j(\xf)=c_2[\xf]_p+[\xf-l]_p 
\com\q
\pl_5(\del A_{5\al})|_{\xf=0,l}=0
\label{vac29}
\end{eqnarray}
where $c_2$ is another free parameter.
Because
${\pl_5}^2j=-2lc_2\del'(\xf-l)-2l\del'(\xf)$, 
the solution $j$ of (\ref{vac29}), by itself,  does not satisfy (\ref{vac28})
on the points $\xf=0,l$. In order to correct it, 
we must require the variation $\del A_{5\al}$,
on the points $\xf=0,l$, to satisfy 
the Neumann boundary condition(the second relation of (\ref{vac29})). 
This condition "absorbs" the singularities appearing
in the variation equation 
(used to derive the field equation)
at the points $\xf=0,l$ 
and makes the "free" wave property (\ref{vac28})
consistent everywhere in the extra space. 

Summarizing the case 3) solution, we have
\begin{eqnarray}
\vp_\al=\vpbar_\al\{ c_1[\xf]_p+[\xf-l]_p\} \com\q
a_{5\al}=\abar_\al\{c_2[\xf]_p+[\xf-l]_p \} \com\nn
\eta=\mbox{const}\com\q\eta'=\mbox{const}\com\q
\abar_\al=c_3\vpbar_\al\com\q
\vpbar_\al=\frac{g}{2l}\eta^\dag T^\al\eta=
\frac{1}{c_1}\frac{g}{2l}\eta'^\dag T^\al\eta'\com\nn
\chi^3_\al=-g\{ \del(\xf)+c_1\del(\xf-l)\}\eta^\dag T^\al\eta\ ,\ 
Z_\al=d_\al
=-\frac{g}{2l}(1+c_1)\eta^\dag T^\al\eta\com\nn
\mbox{with the boundary condition: }\pl_5(\del A_{5\al})|_{\xf=0,l}=0
\com
\label{vac30}
\end{eqnarray}
where $c_1, c_2$ and $c_3$ are three free parameters.
The meaning of $c_1$ is the scale freedom in the "parallel"
condition of brane sources 
$\eta'^\dag T^\al\eta'\propto \eta^\dag T^\al\eta$
,
and that of $c_2$ and $c_3$ is the "free" wave property
of $a_{5\al}$ 
.
The bulk {\it scalar}
configuration influences the boundary source fields
through the parameter $c_1$, whereas the bulk {\it vector}
(5th component) does not have such effect. Instead
the latter one satisfies the field equation 
only within the restricted variation (Neumann boundary condition).

This solution includes the cases 1) and 2)
as described below. Some special cases are listed as follows.\nl
\nl
(3A) $c_3=0$\nl
This is the case 1). 
There are some special cases.

3A-a)\ $c_1=-1$\nl
$\eta=\mbox{const}\com\q\eta'=\mbox{const}\ ,\ 
\eta'^\dag T^\al\eta'=-\,\eta^\dag T^\al\eta\ ,$\nl
$a_{5\al}=0\ ,\ 
\vp_\al=-\frac{g}{2}\eta^\dag T^\al\eta\,\ep(\xf)\ ,\ 
\chi^3_\al=-g(\del(\xf)-\del(\xf-l))\eta^\dag T^\al\eta\ ,$\nl
$Z_\al=d_\al=0$\nl

3A-b)\ $c_1=0$\nl
$\eta=\mbox{const}\com\q\eta'^\dag T^\al\eta'=0\ ,$\nl
$a_{5\al}=0\ ,\ 
\vp_\al=\frac{g}{2l}\eta^\dag T^\al\eta [\xf-l]_p\ ,\ 
\chi^3_\al=-g\del(\xf)\eta^\dag T^\al\eta\ ,$\nl
$Z_\al=d_\al=-\frac{g}{2l}\eta^\dag T^\al\eta$\nl

3A-c)\ $1/c_1\ra 0$\nl
$\eta'=\mbox{const}\com\q\eta^\dag T^\al\eta=0\ ,$\nl
$a_{5\al}=0\ ,\ 
\vp_\al=\frac{g}{2l}\eta'^\dag T^\al\eta' [\xf]_p\ ,\ 
\chi^3_\al=-g\del(\xf-l)\eta'^\dag T^\al\eta'\ ,$\nl
$Z_\al=d_\al=-\frac{g}{2l}\eta'^\dag T^\al\eta'$
\nl

3A-b) and 3A-c) are symmetric under the brane and anti-brane
exchange. 3A-a) is self (anti)symmetric.
\nl
\nl
(3B) $c_1=-1$\nl
This is the case 2).\nl
\nl
(3C) $c_1=0$\nl
\nl
(3D) $1/c_1\ra 0$\nl
\nl
(3E) $c_2=-1$\nl
This is the case where the roles of the extra component
of the bulk vector and the bulk scalar are exchanged
in Case (3B). 
\nl
\nl
(3F) $c_2=0$\nl
\nl
(3G) $1/c_2\ra 0$\nl

Another special cases are given by fixing two
parameters, $c_1$ and $c_2$ (keeping the $c_3$-freedom),
as shown in Table 1.

$$
\begin{array}{c|c|c|c}
             &  \begin{array}{c} c_2=-1\\ \pl_5(\del A_{5\al})|_{\xf=0,l}=0\end{array}      
                        & \begin{array}{c} c_2=0\\ \pl_5(\del A_{5\al})|_{\xf=0}=0\end{array}    
                                       &  \begin{array}{c} c_2=1/0\\ \pl_5(\del A_{5\al})|_{\xf=l}=0\end{array} \\
\hline
 & & & \\
\begin{array}{c} c_1=-1 \\ (\eta'^\dag T^\al\eta' \\ \ =-\eta^\dag T^\al\eta)  \end{array}   & 
\begin{array}{c}
\pl_5\vp_\al:\ B\Bbar\\
\pl_5a_{5\al}:\ B\Bbar\\
d_\al=0
\end{array}
          &
\begin{array}{c}
\pl_5\vp_\al:\ B\Bbar\\
\pl_5a_{5\al}:\ B\\
d_\al=0
\end{array}
             &
\begin{array}{c}
\pl_5\vp_\al:\ B\Bbar\\
\pl_5a_{5\al}:\ \Bbar\\
d_\al=0
\end{array}
                                     \\
\hline
 & & & \\
\begin{array}{c} c_1=0 \\ (\eta'^\dag T^\al\eta'=0) \end{array}  &
\begin{array}{c}
\pl_5\vp_\al:\ B\\
\pl_5a_{5\al}:\ B\Bbar\\
d_\al=-\frac{g}{2l}\eta^\dag T^\al\eta
\end{array}
          &
\begin{array}{c}
\pl_5\vp_\al:\ B\\
\pl_5a_{5\al}:\ B\\
d_\al=-\frac{g}{2l}\eta^\dag T^\al\eta
\end{array}
               &
\begin{array}{c}
\pl_5\vp_\al:\ B\\
\pl_5a_{5\al}:\ \Bbar\\
d_\al=-\frac{g}{2l}\eta^\dag T^\al\eta
\end{array}
                                         \\
\hline
 & & & \\
\begin{array}{c} c_1=1/0 \\ (\eta^\dag T^\al\eta=0) \end{array}   &
  \begin{array}{c}
\pl_5\vp_\al:\ \Bbar\\
\pl_5a_{5\al}:\ B\Bbar\\
d_\al=-\frac{g}{2l}\eta'^\dag T^\al\eta'
\end{array}
            & 
\begin{array}{c}
\pl_5\vp_\al:\ \Bbar\\
\pl_5a_{5\al}:\ B\\
d_\al=-\frac{g}{2l}\eta'^\dag T^\al\eta'
\end{array}
                &
\begin{array}{c}
\pl_5\vp_\al:\ \Bbar\\
\pl_5a_{5\al}:\ \Bbar\\
d_\al=-\frac{g}{2l}\eta'^\dag T^\al\eta'
\end{array}
                                            \\
\multicolumn{4}{c}{\q}                        \\
\multicolumn{4}{c}{
\mbox{
Table\ 1\ \ Various vacuum configurations of
the Mirabelli-Peskin model.
      } 
                  }           \\
\multicolumn{4}{c}{
\mbox{
\qqq $B\Bbar$, $\Bbar$ and $B$ correspond to brane-anti-brane, anti-brane
and brane 
      } 
                  }            \\
\multicolumn{4}{c}{
\mbox{
\qqq respectively. See Fig.3.
      } 
                  }
\end{array}
$$
We have solved only (\ref{vac1a}), (\ref{vac1b}) and (\ref{vac1c}). 
When $m_{\alp\bep}$ and $\la_{\alp\bep\gap}$ are given,
the equations (\ref{vac2a}) and (\ref{vac2b}) should be
furthermore solved for $\eta, \eta', f$ and $f'$
using the obtained result. 
The solutions in the second row ($c_1=-1$) of Table 1 correspond
to the SUSY invariant vacuum, irrespective of
whether the vacuum expectation values of the brane-matter
fields ($\eta$ and $\eta'$) vanish or not. 
For other solutions, however, $d_\al$ depends on
$\eta$ or $\eta'$, hence the SUSY symmetry of the vacuum
is determined by the brane-matter fields. 
The eqs. (\ref{vac2a}) and (\ref{vac2b}) have a 'trivial' solution
$\eta=0, f=0$ (or $\eta'=0, f'=0$) when 
$d_\al=-\frac{g}{2l}\eta^\dag T^\al\eta$ 
(or $d_\al=-\frac{g}{2l}{\eta'}^\dag T^\al\eta'$). 
It corresponds to the SUSY invariant vacuum. 
If the equations have a solution $\eta\neq 0$ (or $\eta'\neq 0$), 
it corresponds to a SUSY-breaking vacuum. 

We see the bulk scalar $\Phi$ is localized on the wall(s) where
the source(s) exists, whereas the extra component of the bulk vector
$A_5$ on the wall(s) where the Neumann boundary condition 
is imposed.
The two cases, ($c_1=-1, c_2=-1$) and ($c_1=1/0, c_2=1/0$), 
are treated in \cite{IM0402}.

{\bf 4}\ {\it Fermion Localization, Stability, and Bulk Higgs Mechanism}\q
The vacuum is basically determined by the scalar fields as explained
so far. 
Let us examine the small fluctuation of bulk fermions (gauginos) 
around the background solution obtained previously.
We take ($c_1=-1, c_2=-1$) solution as a representative one. We assume
$\eta\neq 0, \eta'\neq 0$. The relevant part of the Lagrangian is
$
-i\labar_i\ga^M\na_M\la^i+g\labar_i[\Phi,\la^i]
$.
We consider a simple case of G=U(1). The field equation for $\la_L$ 
is given by
\begin{eqnarray}
-i\{\ga^m\pl_m\la_L+\ga^5\pl_5\la_L-g a_5(\xf)\ga^5\la_L\}
+ig\vp(\xf)\la_L=0\com\nn
\vp(\xf)=-\frac{g}{2}\eta^\dag\eta\,\ep(\xf)\com\q
a_5(\xf)=c_3\vp(\xf)
\pr
\label{higgs-1}
\end{eqnarray}
(The same thing can be said for $\la_R$.)
We focus on the fermion zero-mode with chirality $\pm 1$:\ 
$
\la_L=\si(x^m)\om(\xf), \ga^m\pl_m\si=0, \ga^5\si=\pm\si
$. Then the extra-space behaviour $\om(\xf)$ is obtained as
\begin{eqnarray}
\om\propto\exp\{ -\frac{g^2}{2}(1\pm c_3)\eta^\dag\eta\, |\xf|\}
\pr
\label{higgs0}
\end{eqnarray}
As far as $1\pm c_3>0$, the fermion zero mode is localized
around the brane. (If we require fermions with both chiralities
to be localized, we must choose the parameter $c_3$ as
$-1<c_3<1$.)


In the present approach, ($\Ncal=1$)SUSY is basically respected. If SUSY is
preserved, the solutions obtained previously
are expected to be stable, because the force between
branes (Casimir force) vanish from the symmetry.
In some cases, we can more strongly confirm the stableness from the topology
(or index) as follows. 
We can regard the extra-space size ($S^1$ radius) $l$ as
an {\it infrared regularization} parameter for
the {\it non-compact} extra-space {\bf R}($-\infty<y<\infty$). 
By letting $l\ra +\infty$ in the previous result,
we can obtain the vacuum solutions in this case.
First we note
\begin{eqnarray}
\ep(y)\ra {\tilde \ep}(y)\com\q
\frac{1}{l}[y]_P\ra \frac{1}{l}y\com\q
\frac{1}{l}[y-l]_P=\frac{1}{l}[y]_P-\ep(y)\ra \frac{1}{l}y-\eptil(y)\com\nn
\frac{d\ep(y)}{dy}\ra 2\deltil(y)\com\q
\frac{1}{l}\frac{d}{dy}[y]_P\ra 0\com\q
\frac{1}{l}\frac{d}{dy}[y-l]_P\ra -2\deltil(y)\com\nn 
\mbox{  as  }l\ra\infty\pr\qqqqq
\label{higgs1}
\end{eqnarray}
An interesting case is the $l\ra\infty$ limit of
($c_1=-1, c_2=-1$) in Table 1. 
\begin{eqnarray}
\vp_\al=-\vpbar_\al l\ep(\xf)\ra 
-\frac{g}{2}\eta^\dag T^\al\eta\eptil(\xf)\com\q
a_{5\al}=-\abar_\al l\ep(\xf)\ra 
-c_3\frac{g}{2}\eta^\dag T^\al\eta\eptil(\xf)\com\nn
\chi^3_\al=-g\del(\xf)\eta^\dag T^\al\eta\com\q Z_\al=d_\al=0\com\nn
\mbox{with the boundary condition: }\pl_5(\del A_{5\al})|_{\xf=0}=0
\pr
\label{higgs2}
\end{eqnarray}
Indeed we can confirm 
the above limit is a solution of 
\begin{eqnarray}
S=\int d^4x\int_{-\infty}^{+\infty}dx^5\{\Lcal_{blk}+
{\tilde \del}(x^5)\Lcal_{bnd}\}
\com\q
-\infty <\xf<\infty\com
\label{higgs3}
\end{eqnarray}
where $\Lcal_{blk}$ and $\Lcal_{bnd}$ are the same as in Sec.2
except that fields are no more periodic. 
The stableness is clear from the same situation as the kink solution
of Sec.1. On the other hand, in the $l\ra\infty$ limit of ($c_1=1/0, c_2=1/0$)
there remains no localization configuration. 

As a bulk Higgs model, which embodies the non-singular treatment
(kink-generalization) of the singular solution, 
(\ref{higgs2}) and (\ref{higgs3}), we can present the following one.
We make use of the $\Ncal=1$ chiral superfield\cite{Hebec01}: 
$
\Si=\Phi+iA_5+\sqtwo\theta (-i\sqtwo\la_R)+\theta^2 (X^1+iX^2)
$,
which appears, along with $\Ncal=1$ vector multiplet,
in the $Z_2$-parity decomposition explained in Sec.2. 
We propose the following model.
\begin{eqnarray}
S=\int_{-\infty}^\infty d\xf\int d^4x
\{\Lcal_{blk}+(P(\Si)|_{\theta^2}+\mbox{h.c.})
+\Lcal_{matter}
\}\com\nn
P(\Si)=m^2\Si-\frac{\la}{3}\Si^3
\com\label{higgs4}
\end{eqnarray}
where $m$ and $\la$ are a mass parameter and a (dimensionless) coupling
constant respectively. 
$\Lcal_{matter}$ is the matter lagrangian made of 
the 5D SUSY hypermultiplet\cite{Hebec01}:\ 
$H^1,H^2$, two complex scalar fields;\ $\Psi$, one Dirac field;\ 
$F_1,F_2$, two auxiliary fields. 
The brane thickness parameter is given by
$m/\sqrt{\la}$ as a vacuum expectation value of $\Phi+iA_5$. 
In this model, the complex scalar field in the chiral multiplet
plays the role of "radion"  although the present "radion" 
determines not the extra-space size ( $l\ra \infty$ in the present model)
but the brane thickness. We expect the above model gives a {\it non-singular} 
brane solution 
(kink solution in the extra-space) which is both stable and supersymmetric.

{\bf 5}\ {\it Conclusion}\q
In the brane system appearing in string/D-brane theory,
the stableness is the most important requirement. 
We find some stable brane configurations in the SUSY
bulk-boundary theory. 
We systematically solve the singular field equation
using a general mathematical result about the free-wave solution
in $S_1/Z_2$-space. 
The two scalars, the extra-component of the bulk-vector
($A_5$) and the bulk-scalar($\Phi$), constitute the solutions. 
Their different roles are clarified. 
The importance of the "parallel" configuration is disclosed. 
The boundary condition (of $A_5$) and the boundary matter fields
are two important elements for making the localized
configuration. 
Among all solutions, the solution ($c_1=-1,c_2=-1$) is expected to be
the thin-wall limit of a kink solution. We present
a bulk Higgs model corresponding to the non-singular solution.
The model is expected to give a non-singular and stable brane solution
in the SUSY bulk-boundary theory.

In ref.\cite{IM0402,IM0312}, the 1-loop effective potential is obtained
for the backgrounds $(c_1=-1, c_2=-1)$. 
In ref.\cite{IM0403}, a bulk effect in the 1-loop effective potential
is analyzed in relation to the SUSY breaking. 
We hope the family of present solutions will be used for
further understanding of the bulk-boundary system.


\vs 1



\begin{thebibliography}{99}
\bibitem{Rajara82} 
R. Rajaraman, {
{\it Solitons and instantons}, North-Holland Pub., Amsterdam,C1982
}
\bibitem{RS9905} 
L.Randall and R.Sundrum, {
\PRL {\bf 83}(1999)3370,hep-ph/9905221
}
\bibitem{RS9906} 
L.Randall and R.Sundrum, {
\PRL {\bf 83}(1999)4690,hep-th/9906064
}
\bibitem{GW9907} 
W. Goldberger and M. Wise, {\PR {\bf D60}(1999)107505;\ 
\PRL {\bf 83}(1999)4922}
\bibitem{SI0003}   
S.Ichinose,{Class.Quant.Grav.{\bf 18}(2001)421,hep-th/0003275 
}
\bibitem{SI0107}   
S.Ichinose,{Class.Quant.Grav.{\bf 18}(2001)5239,hep-th/0107254
}
\bibitem{HW96} 
P.Ho\v{r}ava and E.Witten,
{\NP{\bf B460}(1996)506,hep-th/9510209}
\bibitem{MP97} 
E.A.Mirabelli and M.E. Peskin,
Phys.Rev.{\bf D58}(1998)065002, hep-th/9712214
\bibitem{Hebec01} 
A. Hebecker, \NP{\bf B632}(2002)101
\bibitem{IM0312} 
S. Ichinose and A. Murayama, hep-th/0401011, US-03-08
"Quantum Dynamics of A Bulk-Boundary System"
\bibitem{IM0402} 
S. Ichinose and A. Murayama, hep-th/0302029, 
Phys.Lett.{\bf B587}(2004)121
\bibitem{IM0403} 
S. Ichinose and A. Murayama, hep-th/0403080, US-03-07,
to be published in Phys.Lett.B. 
"A Bulk Effect to SUSY Effective Potential in a 5D Super-Yang-Mills Model"
\end{thebibliography}
\end{document}